\documentclass[a4paper,12pt]{article}

\usepackage{amsmath,amssymb}
\usepackage[dvips]{graphicx}

\newcommand{\del}{\partial}
\newcommand{\Xdot}{\dot{X}}

\newcommand{\alphatilde}{\tilde{\alpha}}
\newcommand{\Gcal}{\mathcal{G}}
\newcommand{\Acal}{\mathcal{A}}
\newcommand{\Bcal}{\mathcal{B}}
\newcommand{\Rcal}{\mathcal{R}}
\newcommand{\Tcal}{\mathcal{T}}

\newcommand{\Tcalbar}{\bar{\mathcal{T}}}
\newcommand{\Jcalbar}{\bar{\mathcal{J}}}
\newcommand{\Nbar}{\bar{N}}
\newcommand{\Op}{\mathcal{O}}
\newcommand{\Nhat}{\hat{N}}
\newcommand{\Mhat}{\hat{M}}
\newcommand{\Gtilde}{\tilde{G}}
\newcommand{\gtilde}{\tilde{g}}
\newcommand{\Btilde}{\tilde{B}}
\newcommand{\xitilde}{\tilde{\xi}}

\begin{document}
\renewcommand{\thefootnote}{\fnsymbol{footnote}}
\begin{titlepage}
\hfill
{\hfill \begin{flushright} 
HIP-2011-12/TH
\end{flushright}
  }

\vspace*{10mm}

\begin{center}
{\LARGE \textbf{
Closed string transport coefficients \\
and the membrane paradigm}}
\vspace*{18mm}

\large{Yuya Sasai} \footnote{E-mail: sasai@mappi.helsinki.fi}

\vspace*{13mm}
\large{{\itshape
Helsinki Institute of Physics and Department of Physics \\
University of Helsinki,  
P.O.Box 64, FIN-00014 Helsinki, Finland \\
}}

\end{center}

\vspace*{20mm}

\begin{abstract}
I discuss a correspondence between a fictitious fluid in the black hole membrane paradigm and highly excited closed string states according to the black hole correspondence principle.
I calculate the membrane transport coefficients of an electric NS-NS 2-charged black hole and  transport coefficients of the highly excited closed string states which possess a Kaluza-Klein number and a winding number. Comparing both the transport coefficients at the correspondence point, I show that, except for the bulk viscosity, the membrane transport coefficients are of the same order as the transport coefficients of the closed string states on the stretched horizon. Also, I show that,  except for the bulk viscosity,  both the dimensionless transport coefficients, which are defined by dividing the transport coefficients by the entropy density, are exactly equal if the central charge is 6.

\end{abstract}

\end{titlepage}

\newpage
\renewcommand{\thefootnote}{\arabic{footnote}}
\setcounter{footnote}{0}

\section{Introduction}
A correspondence between a black hole and string states has been pointed out in many works. In \cite{Susskind:1993ws}, it has been proposed that the Bekenstein-Hawking entropy of a macroscopic Schwarzshild black hole could be explained by highly excited neutral string states which cover the stretched horizon of the black hole. 
This correspondence has been generalized in the case of  charged black holes \cite{Sen:1995in, Horowitz:1996nw}.
 In fact,  if the mass and charges of the black hole are set to be equal to those of the string states at the correspondence point $r_H\sim l_s$, the Bekenstein-Hawking entropy agrees with the entropy of the string states \cite{Horowitz:1996nw}.
Although the correct numerical coefficient of the Bekenstein-Hawking entropy could not be reproduced in \cite{Susskind:1993ws, Horowitz:1996nw}, an exact value of the Bekenstein-Hawking entropy of a particular kind of black hole has been derived from string theory in \cite{Strominger:1996sh, Callan:1996dv}. 

So far, the relations between black hole physics and string theory have been investigated mainly by examining their entropies. However, in our previous work \cite{Sasai:2010pz}, we have  proposed that the black hole membrane paradigm \cite{Damour:1978cg, Znajek:1978, Thorne:1986iy, Parikh:1997ma} could also be explained by highly excited string states. 
If we apply a time-dependent homogeneous background metric perturbation to the mixed states of the string, a viscous stress tensor of the string states occurs.
Thus, the string states can be seen as a viscous matter.
We have calculated the shear viscosity of the highly excited string states\footnote{The shear  of the string  is caused by the fluctuation of the off-diagonal spatial component of the background metric, not that of the world sheet metric. In fact, there is no off-diagonal spatial component in the world sheet metric.} and have shown that the shear viscosity of the fictitious fluid in the membrane paradigm is of the same order as that of the highly excited string states on the stretched horizon of the black hole \cite{Sasai:2010pz}. However, in our previous work, we have not examined the correspondence between the other transport coefficients of the black hole and the string states. Also, we have not considered the possible charges of the black hole and string states. In this paper, I generalize the previous settings and check whether the membrane transport coefficients of the black hole agree with the transport coefficients of the highly excited string states when the charges are included.
 
This paper is organized as follows. In section \ref{sec:membrane}, I obtain the membrane transport coefficients of an electric NS-NS 2-charged black hole in $D$ dimensional bosonic closed string backgrounds whose $D-d$ spatial directions are  toroidally compactified. After the short reviews of the $d$ dimensional low energy effective field theory of the massless bosonic closed strings  and the electric NS-NS 2-charged black hole solution, I calculate the membrane transport coefficients induced by the external fields in the effective field theory.
In section \ref{sec:closedstringtrans}, I calculate the transport coefficients of the highly excited closed string states which possess a Kaluza-Klein number and a winding number in a compact direction by using the Kubo's formula. 
In section \ref{sec:bs}, I compare  the membrane transport coefficients of the charged black hole with the closed string transport coefficients according to the black hole correspondence principle \cite{Horowitz:1996nw}. 
In order to compare both the transport coefficients consistently, I assume that the highly excited closed string states are on the stretched horizon of the corresponding black hole because for a distant observer, the physical degrees of freedom of the black hole seem to live near the  horizon.
Then, I show that, except for the bulk viscosity, the membrane transport coefficients are of the same order as the closed string transport coefficients. Also, I show that  except for the bulk viscosity, both the dimensionless transport coefficients, which are defined by dividing the transport coefficients by the entropy density, are exactly equal if the central charge is $6$.
The final section is devoted to the summary and comments. In the Appendix \ref{app}, I calculate the bulk viscosity of the fundamental string states and see the discrepancy with  the membrane paradigm.

\section{Membrane transport coefficients in bosonic closed string backgrounds} \label{sec:membrane}
\subsection{Effective field theory in toroidal compactification}
I start with the low energy effective field theory of the bosonic closed   strings in $D$ dimensional spacetime. The action is given by \cite{Polchinski:1998}
\begin{align}
S=\frac{1}{2\kappa^2_D}\int d^Dx \sqrt{-\Gcal}e^{-2\Phi_D}&\bigg[\Rcal-\frac{1}{12}\Gcal^{MM'}\Gcal^{NN'}\Gcal^{PP'}H_{MNP}H_{M'N'P'} \notag \\
&+4\Gcal^{MN}\del_M\Phi_D\del_N\Phi_D\bigg],
\end{align}
where $\kappa_D$ is a constant, $\mathcal{R}$ is the Ricci scalar associated with the $D$ dimensional metric $\Gcal_{MN}$, $H_{MNP}=\del_MB_{NP}+\del_NB_{PM}+\del_PB_{MN}$, $B_{MN}$ and $\Phi_D$ ($M,N,P=0,\cdots, D-1$) are the antisymmetric tensor and dilaton, respectively.  

Let us consider the toroidal compactification of $D-d$ spatial directions,
\begin{align}
x^a\sim x^a+2\pi R~~~~(a=d, \cdots, D-1),
\end{align}
where $R$ is the compactification radius. Then, the $D$ dimensional metric is parameterized as  \cite{Polchinski:1998}
\begin{align}
\Gcal_{MN}dx^Mdx^N=G_{\mu\nu}dx^{\mu}dx^{\nu}+G_{ab}(dx^a+A^a_{\mu}dx^{\mu})(dx^b+A^b_{\nu}dx^{\nu}), \label{eq:metricdecomp}
\end{align}
where $\mu,\nu=0,\cdots, d-1$, $G_{\mu\nu}$, $A^a_{\mu}$ and $G_{ab}$ are the $d$ dimensional metric, Kaluza-Klein $U(1)$ gauge fields and scalar fields, respectively. Since at energies below $R^{-1}$, one can neglect the extra coordinate dependences of the fields, the action becomes \cite{Polchinski:1998}
\begin{align}
S&=\frac{1}{2\kappa_d^2}\int d^dx \sqrt{-G}e^{-2\Phi_d}[R_G+4G^{\mu\nu}\del_{\mu}\Phi_d \del_{\nu}\Phi_d \notag \\
&-\frac{1}{4}G^{ac}G^{bd}G^{\mu\nu}(\del_{\mu}G_{ab}\del_{\nu}G_{cd}+\del_{\mu}B_{ab}\del_{\nu}B_{cd}) \notag \\
&-\frac{1}{4}G^{\mu\mu'}G^{\nu\nu'}(G_{ab}F^a_{\mu\nu}F^b_{\mu'\nu'}+G^{ab}H_{a\mu\nu}H_{b\mu'\nu'})-\frac{1}{12}G^{\mu\mu'}G^{\nu\nu'}G^{\rho\rho'}\mathcal{H}_{\mu\nu\rho}\mathcal{H}_{\mu'\nu'\rho'}], \label{eq:lowenergyaction}
\end{align}
where $\kappa_d^2=\kappa_D^2/(2\pi R)^{D-d}$, $R_G$ is the Ricci scalar associated with  $G_{\mu\nu}$, $\Phi_d=\Phi_D-\frac{1}{4}\ln \det G_{ab}$ is the $d$ dimensional dilaton and 
\begin{align}
F^a_{\mu\nu}&=\del_{\mu}A^a_{\nu}-\del_{\nu}A^a_{\mu}, \\
H_{a\mu\nu}&=\del_{\mu}B_{\nu a}-\del_{\nu}B_{\mu a}, \\
\mathcal{H}_{\mu\nu\rho}&=(\del_{\mu}B_{\nu\rho}-A^a_{\mu}H_{a\nu\rho})+\text{cyclic permutations of }\mu, \nu, \rho.
\end{align}

To obtain the  Einstein-Hilbert action, we redefine the $d$ dimensional metric and dilaton by
\begin{align}
g_{\mu\nu}&=e^{-\frac{4\Phi}{d-2}}G_{\mu\nu}, \label{eq:Einsteinmetric} \\
\Phi&=\Phi_d-\Phi_{d}^0,
\end{align}
where $\Phi_{d}^0$ is the expectation value of $\Phi_d$. Then, the action (\ref{eq:lowenergyaction}) becomes
\begin{align}
S&=\frac{1}{16\pi G_N}\int d^dx \sqrt{-g}\bigg[R_g-\frac{4}{d-2}\del_{\mu}\Phi \del^{\mu}\Phi \notag \\
&-\frac{1}{4}G^{ac}G^{bd}(\del_{\mu}G_{ab}\del^{\mu}G_{cd}+\del_{\mu}B_{ab}\del^{\mu}B_{cd}) \notag \\
&-\frac{1}{4}e^{-\frac{4\Phi}{d-2}}(G_{ab}F^a_{\mu\nu}F^{b\mu\nu}+G^{ab}H_{a\mu\nu}H_{b}{}^{\mu\nu})-\frac{1}{12}e^{-\frac{8\Phi}{d-2}}\mathcal{H}_{\mu\nu\rho}\mathcal{H}^{\mu\nu\rho} \bigg], \label{eq:effectiveaction}
\end{align}
where $G_N=\kappa_d^2 e^{2\Phi_{d}^{0}}/8\pi$ is the $d$ dimensional Newton constant, $R_g$ is the Ricci scalar associated with $g_{\mu\nu}$ and the Greek indices have been raised with $g_{\mu\nu}$.

\subsection{Black hole solutions}
Let us consider the black hole solutions of (\ref{eq:effectiveaction}). The simplest solution is a Schwarzshild black hole in $d$ dimensional spacetime,
\begin{align}
ds^2=-(1-k(r))dt^2+(1-k(r))^{-1}dr^2+r^2 d\Omega_{d-2}^2, 
\end{align}
where
\begin{align}
k(r)=\bigg(\frac{r_H}{r}\bigg)^{d-3},
\end{align}
and $r_H$ is the horizon radius of the black hole. 

One can construct an electric NS-NS charged black hole from the Schwarzshild solution by using the solution generating method \cite{Horowitz:1996nw, Horowitz:1992jp, Peet:1997es, Peet:2000hn}. Let us review the solution generating method. We consider a black string solution in $d+1$ dimensional spacetime, which can be obtained by simply taking a direct product of the Schwarzshild black hole with the real line $\mathbf{R}$. The metric of the black string is given by
\begin{align}
ds^2=-(1-k(r))dt^2+dz^2+(1-k(r))^{-1}dr^2+r^2 d\Omega_{d-2}^2,
\end{align}
where $z\equiv x^d$ denotes the spatial direction along the black string. Applying a Lorentz boost along $z$ direction,
\begin{align}
t\to t\cosh \alpha_w +z \sinh \alpha_w, \notag \\
z\to z\cosh \alpha_w +t \sinh \alpha_w,
\end{align}
where $\alpha_w\in \mathbf{R}$ and the T-duality transformation \cite{Horowitz:1992jp},
\begin{align}
&\gtilde_{zz}=\frac{1}{g_{zz}},~~~~~\gtilde_{z\mu}=\frac{B_{z\mu}}{g_{zz}}, \notag \\
&\gtilde_{\mu\nu}=g_{\mu\nu}-\frac{g_{z\mu}g_{z\nu}-B_{z\mu}B_{z\nu}}{g_{zz}}, \notag \\
&\Btilde_{z\mu}=\frac{g_{z\mu}}{g_{zz}}, ~~~\Btilde_{\mu\nu}=B_{\mu\nu}-\frac{g_{z\mu}B_{\nu z}-g_{z\nu}B_{\mu z}}{g_{zz}}, \notag \\
&\tilde{\Phi}=\Phi-\frac{1}{2}\ln g_{zz},
\end{align}
we obtain the following black string solution with an electric NS-NS charge,
\begin{align}
ds^2_s&=-\frac{1-k(r)}{f_w(r)}dt^2+\frac{dz^2}{f_w(r)}+(1-k(r))^{-1}dr^2+r^2d\Omega_{d-2}^2, \notag \\
B_{tz}&=\frac{k(r)\cosh \alpha_w \sinh \alpha_w}{f_w(r)}, \notag \\
e^{-2\Phi}&=f_w(r),
\end{align}
where 
\begin{align}
f_w(r)=1+k(r)\sinh^2 \alpha_w.
\end{align}
Here, the metric is expressed in string frame. Applying a Lorentz boost along $z$ direction again,
\begin{align}
t\to t\cosh \alpha_p +z \sinh \alpha_p, \notag \\
z\to z\cosh \alpha_p +t \sinh \alpha_p,
\end{align}
where $\alpha_p \in \mathbf{R}$ and compactifying the $z$ direction on a circle, we obtain an electric NS-NS 2-charged black hole solution of (\ref{eq:effectiveaction}),
\begin{align}
&ds^2_s=-\frac{1-k(r)}{f_w(r)f_p(r)}dt^2+\frac{dr^2}{1-k(r)}+r^2 d\Omega_{d-2}^2, \label{eq:ns2charge} \\
&G_{dd}=\frac{f_p(r)}{f_w(r)},  ~~~~~~~G_{ab}=\delta _{ab} ~~(a,b\neq d), \\
&e^{-4\Phi}=f_w(r)f_p(r),  \\
&A^d_t=\frac{k(r)\cosh \alpha_p \sinh \alpha_p}{f_p(r)},  ~~~~A^a_{\mu}=0~~(a\neq d), \\
&B_{td}=\frac{k(r)\cosh \alpha_w \sinh \alpha_w}{f_w(r)}, ~~~B_{\mu a}=0~~(a\neq d), \label{eq:2chargeb}
\end{align} 
where 
\begin{align}
f_{p}(r)&=1+k(r)\sinh^2\alpha_{p},
\end{align}
and the horizon radius is $r=r_H$. The Einstein metric of this solution is 
\begin{align}
&ds^2=e^{-\frac{4\Phi}{d-2}}ds^2_s \notag \\
&=-(f_w(r)f_p(r))^{-\frac{d-3}{d-2}}(1-k(r))dt^2+\frac{(f_w(r)f_p(r))^{\frac{1}{d-2}}}{1-k(r)}dr^2+(f_w(r)f_p(r))^{\frac{1}{d-2}}r^2d\Omega_{d-2}^2. \label{eq:einsteinchargedbh}
\end{align}
The Arnowitt-Deser-Misner (ADM) mass and electric NS-NS charges are \cite{Peet:1997es}
\begin{align}
M_{BH}&=\frac{(d-3)\omega_{d-2}r_H^{d-3}}{16\pi G_N}\bigg[\frac{1}{d-3}+\frac{1}{2}(\cosh (2\alpha_p)+\cosh (2\alpha_w)) \bigg], \\
q_{p,w}&=\frac{(d-3)\omega_{d-2}r_H^{d-3}}{16\pi G_N}\bigg[\frac{1}{2}\sinh (2\alpha_{p,w}) \bigg],
\end{align}
where $\omega_n=2\pi^{(n+1)/2}/\Gamma(\frac{n+1}{2})$ is the volume of a unit $n$ dimensional sphere.
In the quantum theory, the following quantities are integer normalized \cite{Horowitz:1996nw, Peet:1997es}:
\begin{align}
Q_p=q_pR, ~~~~Q_w=\frac{q_w\alpha'}{R}.
\end{align}
Since the area of the horizon is
\begin{align}
A_{H}=\omega_{d-2}r_H^{d-2}\cosh \alpha_p \cosh \alpha_w, \label{eq:area}
\end{align}
the Bekenstein-Hawking entropy and entropy density are
\begin{align}
S_{BH}&=\frac{A_{H}}{4G_{N}}=\frac{\omega_{d-2}r_H^{d-2}}{4G_N}\cosh \alpha_p \cosh \alpha_w, \label{eq:blackentropy} \\
s_{BH}&=\frac{S_{BH}}{A_{H}}=\frac{1}{4G_N}.
\end{align}

\subsection{Stretched horizon} \label{sec:stretched}
The stretched horizon is defined by the place where the local Unruh temperature becomes the Hagedorn temperature $T\sim 1/l_s$ \cite{Susskind:1993ws}. To find the location of the stretched horizon, we consider the near horizon geometry of the black hole solution (\ref{eq:ns2charge}). Let us define a new radial coordinate $\rho$,
\begin{align}
r^{d-3}-r_H^{d-3}=\frac{(d-3)^2}{4}r_H^{d-5}\rho^2.
\end{align}
Taking the near horizon limit $r\to r_H$, one finds
\begin{align}
ds_s^2\sim -\kappa^2 \rho^2 dt^2 +d\rho^2 +r_H^2d\Omega_{d-2}^2,
\end{align}
where
\begin{align}
\kappa=\frac{d-3}{2r_H}\frac{1}{\cosh \alpha_p \cosh \alpha_w}
\end{align}
is  the surface gravity. Thus, the Hawking temperature of this black hole is 
\begin{align}
T_H=\frac{\kappa}{2\pi}=\frac{d-3}{4\pi r_H \cosh \alpha_p \cosh \alpha_w}.
\end{align}
Since the local Unruh temperature at $\rho$ is given by
\begin{align}
T_{\rho}=\frac{1}{2\pi \rho},
\end{align}
the stretched horizon is located at $\rho\sim l_s$ \cite{Susskind:1993ws}.


\subsection{Membrane transport coefficients induced by scalar fields}
The membrane paradigm states that an observer outside an event horizon of a black hole sees a fictitious viscous and conductive membrane or fluid on the stretched horizon \cite{Damour:1978cg, Znajek:1978, Thorne:1986iy}.
It is known that the shear viscosity of the fictitious membrane is \cite{Thorne:1986iy, Parikh:1997ma, Iqbal:2008by}
\begin{align}
\eta_{BH}=\frac{1}{16\pi G_N}.
\end{align}
This value does not change even if the black hole has some charges. Here, I calculate the other membrane transport coefficients of the electric NS-NS 2-charged black hole.  

At first, I calculate the membrane transport coefficient induced by the dilaton \cite{Parikh:1997ma, Iqbal:2008by}. By varying the action (\ref{eq:effectiveaction}) with respect to the dilation, one finds
\begin{align}
\delta S&=-\frac{4}{d-2}\frac{2}{16\pi G_N}\int d^dx \sqrt{-g}[\nabla_{\mu}\delta \Phi \nabla^{\mu}\Phi +\delta \Phi (\cdots )] \notag \\
&=-\frac{8}{d-2}\frac{1}{16\pi G_N}\int d^dx \sqrt{-g}[\nabla_{\mu}(\delta \Phi \nabla^{\mu}\Phi)-\delta \Phi (\nabla^{\mu} \nabla_{\mu}\Phi +\cdots)], \notag \\
\end{align}
where $\nabla_{\mu}$ is the covariant derivative associated with $g_{\mu\nu}$ and $\cdots$ denotes the other terms of the variation. For an observer outside an event horizon, fields inside the horizon can not possess any classical relevance. Assuming the Dirichlet boundary condition at $r=\infty$, one finds \cite{Parikh:1997ma}
\begin{align}
\delta S=\frac{8}{d-2}\frac{1}{16\pi G_N}\bigg[\int_{\Sigma} d^{d-1}x \sqrt{-h}\delta \Phi n^{\mu}\nabla_{\mu}\Phi+\int d^dx \sqrt{-g}\delta \Phi (\nabla^{\mu} \nabla_{\mu}\Phi+\cdots )\bigg], \label{eq:dvariation2}
\end{align}
where $\Sigma$ denotes the stretched horizon, $h_{\mu\nu}$ is the induced metric on $\Sigma$ and $n^{\mu}$ is the outward-pointing spacelike unit normal to $\Sigma$. To cancel the boundary term of (\ref{eq:dvariation2}),  the following surface term should be added to the action:
\begin{align}
S_{surf}[\Phi]=\int_{\Sigma} d^{d-1}x \sqrt{-h}J_{\Phi}\Phi.
\end{align}
Then, we find
\begin{align}
J_{\Phi}=-\frac{8}{d-2}\frac{1}{16\pi G_N}n^{\mu}\nabla_{\mu}\Phi. \label{eq:dilatoncharge}
\end{align}
$J_{\Phi}$ is interpreted as a charge density on the stretched horizon induced by the bulk dilaton field. 
Since the Einstein metric of the black hole solution (\ref{eq:einsteinchargedbh}) takes the following form,
\begin{align}
ds^2=-g_{tt}(r)dt^2+g_{rr}(r)dr^2+r^2f(r)d\Omega_{d-2}^2, \label{eq:genblackbg}
\end{align} 
the membrane charge density becomes
\begin{align}
J_{\Phi}= -\frac{8}{d-2}\frac{1}{16\pi G_N}\frac{1}{\sqrt{g_{rr}}}\del_r \Phi \big|_{\Sigma}.
\end{align}
Fields measured by a free falling observer must be regular at an event horizon \cite{Thorne:1986iy, Parikh:1997ma}. This is equivalent to the fact that the fields at the event horizon depend only on the ingoing null coordinate $v$ defined by \cite{Iqbal:2008by}
\begin{align}
dv=dt+\sqrt{\frac{g_{rr}}{g_{tt}}}dr.
\end{align}
Thus, near the horizon, we find
\begin{align}
\del_r\Phi\simeq \sqrt{\frac{g_{rr}}{g_{tt}}}\del_t\Phi.
\end{align}
Therefore, the membrane charge density becomes
\begin{align}
J_{\Phi}\simeq -\frac{8}{d-2}\frac{1}{16\pi G_N}\frac{1}{\sqrt{g_{tt}}}\del_t \Phi \big|_{\Sigma}=-\frac{8}{d-2}\frac{1}{16\pi G_N}U^{\mu}\del_{\mu}\Phi,
\end{align}
where $U^{\mu}$ is the velocity vector of an observer at the stretched horizon. Thus, the membrane transport coefficient induced by the dilaton field is 
\begin{align}
\chi_{\Phi}=\frac{8}{d-2}\frac{1}{16\pi G_N}.
\end{align}

In the same way, one can calculate the membrane transport coefficients induced by the scalar fields $G_{ab}$ and $B_{ab}$. Let us consider the case of $G_{ab}$. The surface term which we have to add to the action takes the following form:
\begin{align}
S_{surf}[G_{ab}]=\frac{1}{2}\int_{\Sigma}d^{d-1}x \sqrt{-h}J_{G}^{ab}G_{ab}.
\end{align}
Then, we find
\begin{align}
J_{G}^{ab}=-\frac{1}{16\pi G_N}G^{ac}G^{bd}n^{\mu}\nabla_{\mu}G_{cd}.
\end{align}
Lowering the indices $a$ and $b$, we  obtain
\begin{align}
J^{G}_{ab}=-\frac{1}{16\pi G_N}n^{\mu}\nabla_{\mu}G_{ab}. \label{eq:jgab}
\end{align}
Comparing (\ref{eq:jgab}) with (\ref{eq:dilatoncharge}), we can read off the membrane transport coefficients induced by $G_{ab}$. Since the membrane transport coefficients induced by $B_{ab}$ can be obtained by the same way, we find
\begin{align}
\chi_{G_{ab}}=\chi_{B_{ab}}=\frac{1}{16\pi G_N}. 
\end{align}

\subsection{Membrane conductivities induced by gauge fields}
There are two kinds of gauge fields, $A^a_{\mu}$ and $B_{\mu a}$.  At first, I consider the case of $A^a_{\mu}$ \cite{Parikh:1997ma, Iqbal:2008by}.
By varying the action with respect to $A^a_{\mu}$, one obtains
\begin{align}
\delta S=\frac{1}{16\pi G_N}\bigg[\int_{\Sigma} d^{d-1}x \sqrt{-h}e^{-\frac{4\Phi}{d-2}}G_{ab}n_{\mu}F^{b\mu\nu}\delta A^a_{\nu}+ (\text{bulk terms})\bigg].
\end{align}
To cancel the boundary term, the following surface term should be added to the action,
\begin{align}
S_{surf}[A^a_{\mu}]=\int_{\Sigma}d^{d-1}x \sqrt{-h}J_{A}{}^{a\mu}A_{a\mu}.
\end{align}
Then, we find 
\begin{align}
J_{A}{}^{a \mu}=-\frac{1}{16\pi G_N}e^{-\frac{4\Phi}{d-2}}n_{\nu}F^{a\nu\mu}.
\end{align}
$J_{A}{}^{a\mu}$ is interpreted as the membrane current induced by the bulk gauge field $A_{a\mu}$. In the electric NS-NS 2-charged black hole background, the membrane current becomes
\begin{align}
J_{A}{}^{a\mu}=-\frac{1}{16\pi G_N}e^{-\frac{4\Phi}{d-2}}\frac{1}{\sqrt{g_{rr}}}F^a{}_{r}{}^{\mu}\big|_{\Sigma}. \label{eq:gaugecurrenthorizon}
\end{align}
Since $A^a_{\mu}$ depends only on the ingoing null coordinate $v$ on the stretched horizon, we find
\begin{align}
J_{A}{}^{ai}\simeq -\frac{1}{16\pi G_N}e^{-\frac{4\Phi}{d-2}}\frac{F^a{}_{t}{}^i}{\sqrt{g_{tt}}}\bigg|_{\Sigma}\simeq \frac{1}{16\pi G_N}e^{-\frac{4\Phi(r_H)}{d-2}}E^{ai},
\end{align}
where $i$ denotes the spatial components orthogonal to the radial direction $r$ and $E^{ai}=-U^{\mu}F^a{}_{\mu}{}^{i}$ is the electric field measured by an observer on the stretched horizon. Thus, the membrane conductivities induced by $A^a_{i}$ are
\begin{align}
\chi_{A^a_{i}}=\frac{1}{16\pi G_N}e^{-\frac{4\Phi(r_H)}{d-2}}. \label{eq:aconductivity}
\end{align}

Next, I consider the case of $B_{\mu a}$. To cancel the boundary term, we have to add the following surface term to the action,
\begin{align}
S_{surf}[B]=\int_{\Sigma_H}d^{d-1}x\sqrt{-h}J_B^{\mu a}B_{\mu a}.
\end{align}
Then, we find
\begin{align}
J_B^{\mu a}=-\frac{1}{16\pi G_N}n_{\nu}(e^{-\frac{4\Phi}{d-2}}H^{a\nu\mu}-e^{-\frac{8\Phi}{d-2}}A^a_{\rho}\mathcal{H}^{\nu\mu\rho}). \label{eq:jbmembrane}
\end{align}
In the electric NS-NS 2-charged black hole background, the second term of (\ref{eq:jbmembrane}) vanishes and we obtain
\begin{align}
J_{B}^{\mu a}=-\frac{1}{16\pi G_N}e^{-\frac{4\Phi}{d-2}}\frac{1}{\sqrt{g_{rr}}}H^a{}_{r}{}^{\mu}\big|_{\Sigma_H}. \label{eq:jbmua}
\end{align}
Comparing (\ref{eq:jbmua}) with (\ref{eq:gaugecurrenthorizon}), the membrane conductivities induced by $B_{i a}$ are found to be
\begin{align}
\chi_{B_{i a}}=\frac{1}{16\pi G_N}e^{-\frac{4\Phi(r_H)}{d-2}}. \label{eq:bconductivity}
\end{align}

\subsection{Membrane conductivity induced by antisymmetric tensor field}
Finally, I calculate the membrane conductivity induced by the antisymmetric tensor field. By varying the action (\ref{eq:effectiveaction}) with respect to $B_{\mu\nu}$, one finds
\begin{align}
\delta S=\frac{1}{32\pi G_N}\bigg[\int_{\Sigma_H} d^{d-1}x \sqrt{-h}e^{-\frac{8\Phi}{d-2}}n_{\mu}H^{\mu\nu\rho}\delta B_{\nu\rho}+(\text{bulk terms})\bigg],
\end{align}
To cancel the boundary term, we add the following surface term,
\begin{align}
S_{surf}[B_{\mu\nu}]=\frac{1}{2}\int_{\Sigma} d^{d-1}x \sqrt{-h}J_B^{\mu\nu}B_{\mu\nu}.
\end{align}
Then, we find
\begin{align}
J_B^{\mu\nu}=-\frac{1}{16\pi G_N}e^{-\frac{8\Phi}{d-2}}n_{\rho}H^{\rho\mu\nu}.
\end{align}
$J_{B}^{\mu\nu}$ is interpreted as the membrane antisymmetric tensor current induced by the bulk antisymmetric tensor field. In the electric NS-NS 2-charged black hole background, the membrane antisymmetric tensor current becomes
\begin{align}
J_B^{\mu\nu}=-\frac{1}{16\pi G_N}e^{-\frac{8\Phi}{d-2}}\frac{1}{\sqrt{g_{rr}}}H_{r}{}^{\mu\nu}\big|_{\Sigma}.
\end{align}
Using the ingoing boundary condition, we obtain
\begin{align}
J_{B}^{ij}\simeq -\frac{1}{16\pi G_N}e^{-\frac{8\Phi}{d-2}}\frac{H_{t}{}^{ij}}{\sqrt{g_{tt}}}\bigg|_{\Sigma}\simeq-\frac{1}{16\pi G_N}e^{-\frac{8\Phi(r_H)}{d-2}}U^{\mu}H_{\mu}{}^{ij}.
\end{align}
Thus, the membrane conductivity induced by $B_{ij}$ is
\begin{align}
\chi_{B_{ij}}=\frac{1}{16\pi G_N}e^{-\frac{8\Phi(r_H)}{d-2}}. \label{eq:btrans}
\end{align}

\section{Closed string transport coefficients} \label{sec:closedstringtrans}
\subsection{Statistical description of closed string states}
A classical black hole solution represents a statistical ensemble of states. Since we would like to see the correspondence between a black hole and a fundamental string, the string should also be  described by a statistical ensemble, namely the mixed states. The statistical equilibrium density matrix of the fundamental closed string is defined by \cite{Halyo:1996xe}
\begin{align}
\rho=Z^{-1}\exp (-\beta_L N_L-\beta_R N_R),
\end{align}
where $Z=tr [\exp (-\beta_L N_L-\beta_R N_R)]$, $N_{L,R}$ are the excitation levels of the closed string and $\beta_{L,R}$ are the conjugate parameters of $N_{L,R}$, respectively. 
Using the density matrix, we can calculate the mean values and fluctuations of the excitation levels and the entropy of the string states as follows \cite{Halyo:1996xe, Damour:1999aw}:
\begin{align}
\bar{N}_{L,R}&\equiv \langle N_{L,R} \rangle=\frac{c\pi^2}{6\beta_{L,R}^2}, \label{eq:averagelevel} \\
(\Delta N_{L,R})^2&\equiv \langle (N_{L,R}-\bar{N}_{L,R})^2 \rangle=\frac{c\pi^2}{3\beta_{L,R}^3}, \\
S=-\langle \ln &\rho \rangle=2\pi \bigg(\sqrt{\frac{c\bar{N}_L}{6}}+\sqrt{\frac{c\bar{N}_R}{6}}\bigg). \label{eq:stringentropy}
\end{align}
Here, $\langle \mathcal{O} \rangle \equiv tr (\rho \mathcal{O})$ and $c=D-2$ is the central charge of the $D$ dimensional bosonic string theory. We have to impose that $\beta_{L,R}$ are much smaller than 1 because when the string states are in statistical equilibrium, the relative fluctuations $\Delta N_{L,R}/\Nbar_{L,R}$ should be much smaller than 1 \cite{Damour:1999aw, Landau}. Thus, in order to admit the statistical description, the string states must be highly excited.

\subsection{Linear response and Kubo's formula}
To obtain the transport coefficients of the highly excited closed string states, I use the Kubo's formula \cite{Sasai:2010pz}. 
Here, I briefly review the linear response theory and the Kubo's formula \cite{Chaikin:book}. Let us consider the following Hamiltonian of a homogeneous system:
\begin{align}
H=H_0-\sum_j \mathcal{B}_j h_j(t),
\end{align}
where $H_0$ is a free Hamiltonian, $h_j(t)$ is a time-dependent classical external field and $\mathcal{B}_j$ is a conjugate operator of $h_j(t)$. Here, $j$ denotes the number of the external fields. The external fields are assumed to be weak and the Hamiltonian is expressed in the Schr\"{o}dinger picture. 

We are interested in the expectation values of physical operators in the presence of the external fields. If the external fields are turned on at a fixed time $t_0$, the expectation value of an operator $\mathcal{A}_i$ at $t~(t>t_0)$ is given by
\begin{align}
\langle \mathcal{A}_i(t) \rangle_h=tr(\rho_h(t, t_0)\mathcal{A}_i(t_0)), \label{eq:expresponse}
\end{align}
where $\rho_h(t, t_0)$ is a time-dependent density matrix in the presence of the external fields. For $t\leq t_0$, $\rho_h$ reduces to a time-independent statistical equilibrium density matrix $\rho$.

Let us find the expression of (\ref{eq:expresponse}) in terms of $\rho$. Since the time dependence of $\rho_h(t)$ is given by
\begin{align}
i\frac{\del \rho_h}{\del t}=[H, \rho],
\end{align}
we find
\begin{align}
\langle \mathcal{A}_i(t) \rangle_h=tr[\rho U^{-1}(t, t_0)\mathcal{A}_i(t) U(t,t_0)], \label{eq:fullresponse}
\end{align}
where
\begin{align}
U(t, t_0)=T \exp\bigg(i \int_{t_0}^{t}dt' \sum_j \mathcal{B}_j(t')h_j(t')\bigg), \label{eq:transt}
\end{align}
$T$ denotes the time-ordered product and the time-dependent operators which appear on the right hand side of (\ref{eq:fullresponse}) are expressed in the interaction picture. 

We set $t_0=-\infty$. Up to the linear order in $h_j(t)$,  (\ref{eq:fullresponse}) becomes
\begin{align}
\delta \langle \mathcal{A}_i(t) \rangle &\equiv \langle \mathcal{A}_i(t) \rangle_h-\langle \mathcal{A}_i(t) \rangle \notag \\
&\simeq \sum_j \int_{-\infty}^{\infty}dt' ~i\theta (t-t')\langle [\mathcal{A}_i(t), \mathcal{B}_j(t')] \rangle h_j(t') \notag \\
&=\sum_j \int_{-\infty}^{\infty}dt' G^R_{\mathcal{A}_{i}\mathcal{B}_{j}}(t-t')h_j(t'), \label{eq:linearres}
\end{align}
where $\theta(t)$ is the Heaviside step function and $\langle \Op \rangle\equiv tr(\rho \Op)$. $G^R_{\mathcal{A}_{i}\mathcal{B}_{j}}(t-t')$ is called the response function. After the Fourier transformation, (\ref{eq:linearres}) becomes
\begin{align}
\delta \Acal_i(\omega)=\sum_j G^{R}_{\Acal_i\Bcal_j}(\omega)h_j(\omega).
\end{align}
It is known that the real part of $G^{R}_{\Acal_i\Bcal_j}(\omega)$ is even in $\omega$ and the imaginary part of $G^{R}_{\Acal_i\Bcal_j}(\omega)$ is odd in $\omega$ \cite{Chaikin:book}. Especially, in the low frequency limit $\omega \to 0$, this expression takes the following form:
\begin{align}
\delta \Acal_i(\omega)\sim \sum_j(\gamma_{\Acal_i\Bcal_j}+i\omega \chi_{\Acal_i\Bcal_j})h_j(\omega),
\end{align}
where $\gamma_{\Acal_i\Bcal_j}$ and $\chi_{\Acal_i\Bcal_j}$ are the leading coefficients of the real part and imaginary of $G^{R}_{\Acal_i\Bcal_j}(\omega)$ in $\omega$, respectively. The dissipative part of the operator $\Acal_i(t)$ is given by the second term and $\chi_{\Acal_i\Bcal_j}$ is called the transport coefficient.

To obtain the transport coefficients, it is convenient to introduce the following function,
\begin{align}
f_{\Acal_i\Bcal_j}(t-t')=\frac{1}{2}\langle [\mathcal{A}_i(t), \mathcal{B}_j(t')] \rangle.
\end{align}
It is known that  \cite{Chaikin:book}
\begin{align}
Im[G^{R}_{\mathcal{A}_i\mathcal{B}_j}(\omega)]=f_{\Acal_i\Bcal_j}(\omega).
\end{align}
Thus, in the low frequency limit $\omega \to 0$, we obtain
\begin{align}
f_{\Acal_i\Bcal_j}(\omega)\simeq \chi_{\Acal_i\Bcal_j}\omega, \label{eq:kubo}
\end{align}
which is known as the Kubo's formula.

\subsection{Closed string theory in  light-cone gauge}
Let us start with the world sheet action of the $D$ dimensional bosonic string theory. The action is given by\footnote{The world sheet metric has been chosen to be flat.}
\begin{align}
S=\frac{1}{4\pi \alpha'}\int d^2 \sigma [\Gcal_{MN}(X)(\dot{X}^{M}\dot{X}^N-X^{'M}X^{'N})+B_{MN}(X)(X^{'M}\dot{X}^{N}-\dot{X}^{M}X^{'N})],
\end{align}
where $\sigma^{\alpha}=(\tau, \sigma) ~(\alpha=0,1)$ are the world sheet coordinates and
\begin{align}
\Xdot^{M}=\frac{\del X^{M}}{\del \tau}, ~~X^{'M}=\frac{\del X^{M}}{\del \sigma}.
\end{align}
If $D-d$ spatial directions are toroidally compactified, the action becomes 
\begin{align}
S=\frac{1}{4\pi \alpha'}\int d^2 \sigma [&(e^{\frac{4\Phi}{d-2}}g_{\mu\nu}+G_{ab}A^a_{\mu}A^b_{\nu})P^{\mu\nu}+2G_{ab}A^b_{\mu}P^{a \mu}+G_{ab}P^{ab} \notag \\
&+B_{\mu\nu}Q^{\mu\nu}+2B_{\mu a}Q^{\mu a}+B_{ab}Q^{ab}], \label{eq:worldsheet2}
\end{align}  
where I have used (\ref{eq:metricdecomp}) and (\ref{eq:Einsteinmetric}) and have defined
\begin{align}
P^{MN}(\tau, \sigma)&=\dot{X}^{M}\dot{X}^N-X^{'M}X^{'N}, \label{eq:pmn} \\
Q^{MN}(\tau, \sigma)&=X^{'M}\dot{X}^{N}-\dot{X}^{M}X^{'N}. \label{eq:qmn}
\end{align}
By definition, $P^{MN}$ and $Q^{MN}$ are symmetric and antisymmetric under the exchange of $M$ and $N$, respectively. Expanding the background fields around the trivial backgrounds, one finds
\begin{align}
S&\simeq S_0+S_1, \\
S_0&=\frac{1}{4\pi \alpha'}\int d^2\sigma \eta_{MN}(\dot{X}^{M}\dot{X}^N-X^{'M}X^{'N}), \label{eq:standard} \\
S_1&=\frac{1}{4\pi \alpha'}\int d^2\sigma [h_{\mu\nu}(X)P^{\mu\nu}+\frac{4}{d-2}\Phi(X) \eta_{\mu\nu}P^{\mu\nu}+\Gtilde_{ab}(X)P^{ab}+B_{ab}(X)Q^{ab} \notag \\
&~~~~~~~~~~~~~~~~~~~+2A_{a\mu}(X)P^{a\mu}+2B_{\mu a}(X)Q^{\mu a}+B_{\mu\nu}(X)Q^{\mu\nu}], \label{eq:sourcestring}
\end{align}
where $\eta_{MN}$ is the flat metric, $\Gtilde_{ab}=G_{ab}-\delta_{ab}$, $S_0$ is the free parts of the action and $S_1$ is composed of the linear terms in the background perturbations. In this section, $h_{\mu\nu}$ denotes the metric perturbation around the flat spacetime.

I consider the closed string  which possess a Kaluza-Klein number $K$ and a winding number $W$ in the direction of $x^d$. The mode expansions of the closed string are given by\footnote{For simplicity, the constant terms in the mode expansions are set to be zero.}
\begin{align}
X^{\mu}(\tau, \sigma)&=2\alpha' p^{\mu}\tau +i\sqrt{\frac{\alpha'}{2}} \sum_{n\neq 0}\bigg[\frac{\alpha^{\mu}_n}{n}e^{-2in(\tau-\sigma)}+\frac{\alphatilde^{\mu}_n}{n}e^{-2in(\tau+\sigma)} \bigg], \label{eq:modeexp1} \\
X^{d}(\tau, \sigma)&=2\alpha' \frac{K}{R}\tau +2WR\sigma +i\sqrt{\frac{\alpha'}{2}} \sum_{n\neq 0}\bigg[\frac{\alpha^{a}_n}{n}e^{-2in(\tau-\sigma)}+\frac{\alphatilde^{a}_n}{n}e^{-2in(\tau+\sigma)} \bigg], \label{eq:modeexp2} \\
X^{a}(\tau, \sigma)&=i\sqrt{\frac{\alpha'}{2}} \sum_{n\neq 0}\bigg[\frac{\alpha^{a}_n}{n}e^{-2in(\tau-\sigma)}+\frac{\alphatilde^{a}_n}{n}e^{-2in(\tau+\sigma)} \bigg]~~~~~~(a\neq d), \label{eq:modeexp3}
\end{align}
where $\alpha_m^{M}$ and $\alphatilde_m^{M}$ satisfy the following commutation relations:
\begin{align}
&[\alpha_m^{M}, \alpha_n^{N}]=m\delta_{m+n,0}\eta^{MN}, \notag \\
&[\alphatilde_m^{M}, \alphatilde_n^{N}]=m\delta_{m+n,0}\eta^{MN}, \notag \\
&[\alpha_m^{M}, \alphatilde_n^{N}]=0. \label{eq:osc}
\end{align}

To solve the Virasoro constraints, we impose the light-cone gauge,
\begin{align}
X^+=2\alpha' p^+\tau, \label{eq:lightconegauge}
\end{align}
where $X^{\pm}=(X^0\pm X^{d-1})/\sqrt{2}$.
Since $X^{+}$ is identified with  $\tau$, the spacetime coordinate $x^{+}$  plays the role of time and $x^{-}$ describes the longitudinal spatial direction. From the Virasoro constraints, 
the mass shell conditions are found to be
\begin{align}
M^2&=(p_L^d)^2+\frac{4N_L}{\alpha'}=(p_R^d)^2+\frac{4N_R}{\alpha'}, \\
p_L^d&=\frac{K}{R}-\frac{WR}{\alpha'}, \\
p_R^d&=\frac{K}{R}+\frac{WR}{\alpha'},
\end{align}
where the excitation levels
\begin{align}
N_L&=\sum_{n=1}^{\infty}:(\alpha^i_{-n}\alpha_{ni}+\alpha^a_{-n}\alpha_{na}):, \\
N_R&=\sum_{n=1}^{\infty}:(\alphatilde^i_{-n}\alphatilde_{ni}+\alphatilde^a_{-n}\alphatilde_{na}):,
\end{align}
are taken to be large and $i=1,\cdots, d-2$ denotes the transverse noncompact spatial directions. The symbol $:~:$ denotes the normal ordering.

Inserting the light-cone gauge into the action, (\ref{eq:sourcestring}) becomes
\begin{align}
S_1&=\frac{1}{4\pi \alpha'}\int d^dx \int d^2\sigma \delta(x^{+}-X^{+})\delta(x^{-}-X^{-})\delta^{d-2}(x^i-X^i) \notag \\
&\times[h_{\mu\nu}(x)P^{\mu\nu}+\frac{4}{d-2}\Phi(x) \eta_{\mu\nu}P^{\mu\nu}+\Gtilde_{ab}(x)P^{ab}+B_{ab}(x)Q^{ab} \notag \\
&~~~+2A_{a\mu}(x)P^{a\mu}+2B_{\mu a}(x)Q^{\mu a}+B_{\mu\nu}(x)Q^{\mu\nu}] \notag \\
&=\frac{1}{8\pi \alpha^{'2}p^+}\int d^dx \int_0^{\pi} d\sigma \delta(x^--X^-)\delta^{d-2}(x^i-X^i) \notag \\
&\times [h_{\mu\nu}(x)P^{\mu\nu}+\frac{4}{d-2}\Phi(x) \eta_{\mu\nu}P^{\mu\nu}+\Gtilde_{ab}(x)P^{ab}+B_{ab}(x)Q^{ab} \notag \\
&~~~+2A_{a\mu}(x)P^{a\mu}+2B_{\mu a}(x)Q^{\mu a}+B_{\mu\nu}(x)Q^{\mu\nu}]\big|_{\tau=\frac{x^+}{2\alpha'p^+}}, \label{eq:s1spacetime}
\end{align}
where in the second step, I have performed $\tau$ integration. Thus, (\ref{eq:s1spacetime}) can be seen as the spacetime action. From (\ref{eq:s1spacetime}), one can read off the operators conjugate to the background perturbations. For example, the stress tensor, which is the conjugate operator of $h_{\mu\nu}$, is\footnote{$\Tcal^{\mu\nu}(x^+, x^{-}, x^i)$ is the spacetime stress tensor, not the world sheet stress tensor which is conjugate of the world sheet metric.}
\begin{align}
\Tcal^{\mu\nu}(x^+, x^{-}, x^i)=\frac{1}{4\pi \alpha^{'2} p^+}\int_0^{\pi}d\sigma \delta(x^--X^-)\delta^{d-2}(x^i-X^i)P^{\mu\nu}(\tau, \sigma)\big|_{\tau=\frac{x^+}{2\alpha'p^+}}.
\end{align}

Let us denote $L$ and $L_{-}$ as the transverse size and longitudinal size of the closed string states, respectively.
To obtain the closed string transport coefficients, I consider the operators which are spatially averaged over the volume of the string configuration,
\begin{align}
\bar{\Op}(x^{+})\equiv \frac{1}{V_{d-1}}\int dx^{-}d^{d-2}x^i \Op (x^+, x^-, x^i),
\end{align}
where $V_{d-1}\sim L_{-}L^{d-2}$ \cite{Sasai:2010pz}. Then, all  operators which I will discuss in this section are
\begin{align}
\Tcalbar^{\mu\nu}(x^+)&=\frac{1}{4\pi\alpha^{'2}p^{+}V_{d-1}}\int_0^{\pi} d\sigma P^{\mu\nu}(\tau, \sigma)\big|_{\tau=\frac{x^+}{2\alpha'p^+}}, \\
\Jcalbar_{\Phi}(x^+)&=\frac{1}{8\pi \alpha^{'2}p^{+}V_{d-1}}\frac{4}{d-2} \eta_{\mu\nu}\int_0^{\pi} d\sigma P^{\mu\nu}(\tau, \sigma)\big|_{\tau=\frac{x^+}{2\alpha'p^+}}, \\
\Jcalbar_{A}^{a \mu}(x^+)&=\frac{1}{4\pi \alpha^{'2}p^+V_{d-1}}\int_0^{\pi} d\sigma P^{a \mu}(\tau, \sigma)\big|_{\tau=\frac{x^+}{2\alpha'p^+}}, \\
\Jcalbar_{G}^{ab}(x^+)&=\frac{1}{4\pi \alpha^{'2}p^+V_{d-1}}\int_0^{\pi} d\sigma P^{ab}(\tau, \sigma)\big|_{\tau=\frac{x^+}{2\alpha'p^+}}, \\
\Jcalbar_B^{\mu\nu}(x^+)&=\frac{1}{4\pi \alpha^{'2}p^+V_{d-1}}\int_0^{\pi} d\sigma Q^{\mu\nu}(\tau, \sigma)\big|_{\tau=\frac{x^+}{2\alpha'p^+}}, \\
\Jcalbar_B^{\mu a}(x^+)&=\frac{1}{4\pi \alpha^{'2}p^+V_{d-1}}\int_0^{\pi} d\sigma Q^{\mu a}(\tau, \sigma)\big|_{\tau=\frac{x^+}{2\alpha'p^+}}, \\
\Jcalbar_B^{ab}(x^+)&=\frac{1}{4\pi \alpha^{'2}p^+V_{d-1}}\int_0^{\pi} d\sigma Q^{ab}(\tau, \sigma)\big|_{\tau=\frac{x^+}{2\alpha'p^+}},
\end{align}
where  $\Jcalbar_{\Phi}$, $\Jcalbar_{A}^{a \mu}$, $\Jcalbar_{G}^{ab}$, $\Jcalbar_B^{\mu\nu}$, $\Jcalbar_B^{\mu a}$ and $\Jcalbar_B^{ab}$ are conjugate operators of $\Phi$, $A_{a\mu}$, $\Gtilde_{ab}$, $B_{\mu\nu}$, $B_{\mu a}$ and $B_{ab}$, respectively.

In the membrane paradigm, there is supposed to be some effective fluid on the stretched horizon. 
Thus, to compare the dynamics of the fundamental string states with the membrane paradigm, 
we have to consider the hydrodynamical limit of the fundamental string states. This means that  the wavelengths of the external fields which are applied to the string states must be much longer than  the sizes of the fundamental string states and the frequencies of the external fields must be much shorter than the characteristic frequency of the fundamental string states.
Therefore, I assume that the background perturbations only depend on $x^{+}$. This  corresponds to the infinite wavelength limit.
Then, (\ref{eq:s1spacetime}) becomes
\begin{align}
S_1&=\frac{1}{8\pi \alpha^{'2}p^+}\int dx^{+} \int_0^{\pi} d\sigma [h_{\mu\nu}(x^+)P^{\mu\nu}+\frac{4}{d-2}\Phi(x^+) \eta_{\mu\nu}P^{\mu\nu}+\Gtilde_{ab}(x^+)P^{ab} \notag \\
&~~~~~~~~~~+B_{ab}(x^+)Q^{ab}+2A_{a\mu}(x^+)P^{a\mu}+2B_{\mu a}(x^+)Q^{\mu a}+B_{\mu\nu}(x^+)Q^{\mu\nu}]\big|_{\tau=\frac{x^+}{2\alpha'p^+}} \notag \\
&=V_{d-1}\int dx^{+}\bigg[\frac{1}{2}\Tcalbar^{\mu\nu}h_{\mu\nu}+\Jcalbar_{\Phi}\Phi+\Jcalbar_A^{a\mu}A_{a\mu}+\frac{1}{2}\Jcalbar_{\Gtilde}^{ab}\Gtilde_{ab} \notag \\
&~~~~~~~~~~~~~~~~~~~~~~~~~~~~~~~~~+\frac{1}{2}\Jcalbar_B^{\mu\nu}B_{\mu\nu}+\Jcalbar_B^{\mu a}B_{\mu a}+\frac{1}{2}\Jcalbar_B^{ab}B_{ab} \bigg]. \label{eq:s1ope}
\end{align}

\subsection{Calculations of closed string transport coefficients} \label{sec:cal}
Let us calculate the closed string transport coefficients in the trivial background fields. For simplicity, I set $p^{i}=0$. Inserting the mode expansions (\ref{eq:modeexp1})-(\ref{eq:modeexp3}) into (\ref{eq:pmn}) and (\ref{eq:qmn}), we find
\begin{align}
\int_0^{\pi}d\sigma P^{\hat{M}\hat{N}}(\tfrac{x^+}{2\alpha' p^+}, \sigma)&=4\pi \alpha' \sum_{n\neq 0}(\alpha_n^{\hat{M}}\alphatilde_n^{\hat{N}}+\alphatilde_n^{\hat{M}}\alpha_n^{\hat{N}})e^{-\frac{2inx^+}{\alpha'p^+}}, \\
\int_0^{\pi}d\sigma ~\eta_{\mu\nu}P^{\mu\nu}(\tfrac{x^+}{2\alpha' p^+}, \sigma)&=4\pi \alpha' \delta_{ij}\sum_{n\neq 0}(\alpha_n^i\alphatilde_{n}^j+\alphatilde_n^i\alpha_n^j)e^{-\frac{2inx^+}{\alpha'p^+}} \notag \\
&=\delta_{ij}\int_0^{\pi}d\sigma P^{ij}, \\
\int_0^{\pi}d\sigma Q^{\hat{M}\hat{N}}(\tfrac{x^+}{2\alpha' p^+}, \sigma)&=4\pi \alpha' \sum_{n\neq 0}(\alphatilde_n^{\hat{M}}\alpha_n^{\hat{N}}-\alpha_n^{\hat{M}}\alphatilde_n^{\hat{N}})e^{-\frac{2inx^+}{\alpha'p^+}}, 
\end{align}
where $\hat{M}, \hat{N}=\{i,a\}$. Here, I have omitted the possible constant terms because those terms do not contribute the calculations of the transport coefficients below. 
Using the following formula \cite{Sasai:2010pz, Damour:1999aw},\footnote{I have introduced the normal ordering  to avoid a divergence of the response functions which comes from the zero point energy \cite{Sasai:2010pz, Damour:1999aw}.}
\begin{align}
\langle : \alpha_m^{\hat{M}}\alpha_n^{\hat{N}} : \rangle =\frac{|n|}{e^{\beta_R |n|}-1}\delta^{\hat{M}\hat{N}}\delta_{n+m,0}, \notag \\
\langle : \alphatilde_m^{\hat{M}}\alphatilde_n^{\hat{N}} : \rangle=\frac{|n|}{e^{\beta_L |n|}-1}\delta^{\hat{M}\hat{N}}\delta_{n+m,0},
\end{align}
we find
\begin{align}
\langle : [\int_0^{\pi}d\sigma &P^{\hat{M}\hat{N}}(\tfrac{x^+}{2\alpha' p^+}, \sigma),\int_0^{\pi}d\sigma' P^{\hat{M}'\hat{N}'}(\tfrac{x^{'+}}{2\alpha' p^+}, \sigma')] : \rangle =\delta_+^{\Mhat\Nhat,\Mhat'\Nhat'}I(x^{+}-x^{'+}), \label{eq:ppcom} \\
\langle : [\int_0^{\pi}d\sigma &Q^{\hat{M}\hat{N}}(\tfrac{x^+}{2\alpha' p^+}, \sigma),\int_0^{\pi}d\sigma' Q^{\hat{M}'\hat{N}'}(\tfrac{x^{'+}}{2\alpha' p^+}, \sigma')] : \rangle =\delta_-^{\Mhat\Nhat,\Mhat'\Nhat'}I(x^{+}-x^{'+}), \label{eq:qqcom}
\end{align}
where $\delta_{\pm}^{\Mhat\Nhat,\Mhat'\Nhat'}\equiv \delta^{\Mhat\Mhat'}\delta^{\Nhat\Nhat'}\pm \delta^{\Mhat\Nhat'}\delta^{\Nhat\Mhat'}$ and
\begin{align}
I(x^{+}-x^{'+})\equiv -4i(4\pi \alpha')^2 \sum_{m=1}^{\infty}m^2\bigg(\frac{1}{e^{\beta_R m}-1}+\frac{1}{e^{\beta_L m}-1}\bigg)\sin \bigg(\frac{2m(x^{+}-x^{'+})}{\alpha' p^+}\bigg).
\end{align}
Here, one has to calculate the commutation relations in (\ref{eq:ppcom}) and (\ref{eq:qqcom}) using (\ref{eq:osc}) before considering the normal ordering. The Fourier transformation of the function $I(x^{+})$ is
\begin{align}
I(\omega)&=\int_{-\infty}^{\infty}dx^{+} I(x^{+})e^{i\omega x^{+}} \notag \\
&=-4i(4\pi \alpha')^2 \sum_{m=1}^{\infty}m^2\bigg(\frac{1}{e^{\beta_R m}-1}+\frac{1}{e^{\beta_L m}-1}\bigg)\frac{\pi}{i}\bigg[\delta \bigg(\omega+\frac{2m}{\alpha'p^{+}}\bigg)-\delta \bigg(\omega-\frac{2m}{\alpha'p^{+}}\bigg)\bigg] \notag \\
&=(4\pi \alpha')^3\bigg(\frac{\alpha' p^{+} \omega}{2}\bigg)^2\frac{p^+}{2}\bigg(\frac{1}{e^{\beta_R \alpha' p^{+} \omega/2}-1}+\frac{1}{e^{\beta_L \alpha' p^{+} \omega/2}-1}\bigg),
\end{align}
where I have assumed $\omega$ to be positive. In the low frequency limit $\omega \to 0$, $I(\omega)$ becomes
\begin{align}
I(\omega)\simeq (4\pi \alpha')^3\frac{\alpha' p^{+2}\omega}{4}\bigg(\frac{1}{\beta_R}+\frac{1}{\beta_L}\bigg)=(4\pi \alpha'^2p^{+})^2\sqrt{\frac{6}{c}}(\sqrt{N_R}+\sqrt{N_L})~ \omega,
\end{align}
where I have used (\ref{eq:averagelevel}).

At first, I calculate the shear viscosity of the closed string states. The shear viscosity can be obtained from the response function of the off-diagonal transverse component of the stress tensor $\Tcalbar_{ij}$.   Using the Kubo's formula (\ref{eq:kubo}), we find
\begin{align}
\eta &= \lim_{\omega \to 0}\frac{V_{d-1}}{\omega}f_{\Tcalbar_{ij}\Tcalbar_{ij}}(\omega) \notag \\
&=\frac{V_{d-1}}{2}\bigg(\frac{1}{4\pi\alpha^{'2}p^{+}V_{d-1}}\bigg)^2 \lim_{\omega \to 0}\frac{I(\omega)}{\omega} \notag \\
&=\frac{1}{2V_{d-1}}\sqrt{\frac{6}{c}}(\sqrt{N_R}+\sqrt{N_L}),
\end{align}
where $V_{d-1}$ in the first line comes from the overall proportionality in (\ref{eq:s1ope}). Dividing the shear viscosity by the entropy density of the closed string states,
\begin{align}
s\equiv \frac{S}{V_{d-1}}=\frac{2\pi}{V_{d-1}} \sqrt{\frac{c}{6}}(\sqrt{N_R}+\sqrt{N_L}),
\end{align}
we obtain the ratio of shear viscosity to entropy density,
\begin{align}
\frac{\eta}{s}=\frac{3}{2\pi c}. \label{eq:etaovers}
\end{align}
In our previous work \cite{Sasai:2010pz}, we have calculated the shear viscosity of the open string states. However, $\eta/s$ does not change even if one considers the closed string states.

Next, I calculate the closed string transport coefficient induced by the dilaton. Using the Kubo's formula (\ref{eq:kubo}), we find
\begin{align}
\xi_{\Phi}&= \lim_{\omega \to 0}\frac{V_{d-1}}{\omega}f_{\Jcalbar_{\Phi}\Jcalbar_{\Phi}}(\omega) \notag \\
&=\frac{V_{d-1}}{2}\bigg(\frac{1}{8\pi \alpha^{'2}p^{+}V_{d-1}}\frac{4}{d-2}\bigg)^2\delta_{ij}\delta_{kl}\delta_{+}^{ij,kl}\lim_{\omega \to 0}\frac{I(\omega)}{\omega} \notag \\
&=(d-2)V_{d-1}\bigg(\frac{1}{8\pi \alpha^{'2}p^{+}V_{d-1}}\frac{4}{d-2}\bigg)^2(4\pi \alpha^{'2}p^{+})^2\sqrt{\frac{6}{c}}(\sqrt{N_R}+\sqrt{N_L}) \notag \\
&=\frac{8}{d-2}\frac{1}{2V_{d-1}}\sqrt{\frac{6}{c}}(\sqrt{N_R}+\sqrt{N_L}). \label{eq:xiphi}
\end{align}

In the same way, one can obtain the other transport coefficients. 
The closed string transport coefficients induced by the scalar fields $\Gtilde_{ab}, B_{ab}$ are
\begin{align}
\xi_{G_{ab}}&=\lim_{\omega\to 0}\frac{V_{d-1}}{\omega}f_{\Jcalbar^G_{ab}\Jcalbar^G_{ab}}(\omega) \notag \\
&=\frac{1}{2V_{d-1}}\sqrt{\frac{6}{c}}(\sqrt{N_R}+\sqrt{N_L}), \\
\xi_{B_{ab}}&=\lim_{\omega\to 0}\frac{V_{d-1}}{\omega}f_{\Jcalbar^B_{ab}\Jcalbar^B_{ab}}(\omega) \notag \\
&=\frac{1}{2V_{d-1}}\sqrt{\frac{6}{c}}(\sqrt{N_R}+\sqrt{N_L}).
\end{align}
The closed string transport coefficients induced by the gauge fields $A^a_{\mu}$ and $B_{\mu a}$ are
\begin{align}
\xi_{A^a_{i}}&=\lim_{\omega\to 0}\frac{V_{d-1}}{\omega}f_{\Jcalbar_A{}^a{}_{i}\Jcalbar_A{}^a{}_{i}}(\omega) \notag \\
&=\frac{1}{2V_{d-1}}\sqrt{\frac{6}{c}}(\sqrt{N_R}+\sqrt{N_L}), \\
\xi_{B_{ia}}&=\lim_{\omega\to 0}\frac{V_{d-1}}{\omega}f_{\Jcalbar^B_{ia}\Jcalbar^B_{ia}}(\omega) \notag \\
&=\frac{1}{2V_{d-1}}\sqrt{\frac{6}{c}}(\sqrt{N_R}+\sqrt{N_L}).
\end{align}
The closed string transport coefficient induced by the antisymmetric tensor is
\begin{align}
\xi_{B_{ij}}&=\lim_{\omega\to 0}\frac{V_{d-1}}{\omega}f_{\Jcalbar^B_{ij}\Jcalbar^B_{ij}}(\omega) \notag \\
&=\frac{1}{2V_{d-1}}\sqrt{\frac{6}{c}}(\sqrt{N_R}+\sqrt{N_L}).
\end{align}

\section{Correspondence between membrane transport coefficients and closed string transport coefficients} \label{sec:bs}
\subsection{The black hole correspondence principle}
The black hole correspondence principle states that as one decreases the string coupling $g_s$, a black hole will turn into string states when the curvature length scale at the horizon of the black hole in string frame is of the order of the string length scale $l_s$ \cite{Horowitz:1996nw}. In fact, it has been shown that when the mass and charges of the black hole are set to be equal to those of the string states at the corresponding point, both entropies are equal except for a numerical coefficient \cite{Horowitz:1996nw}.

Let us review the correspondence between the electric NS-NS 2-charged black hole and the highly excited closed string states. Since the curvature at the horizon in string frame is of the order of $r_H^{-2}$, the correspondence point is found to be
\begin{align}
r_H\sim l_s. \label{eq:correspondencepoint}
\end{align}
The Kaluza-Klein number $K$ and the winding number $W$ of the closed string states correspond to the electric charges of the black hole $Q_p$ and $Q_w$, respectively. 
Equating the mass and charges of the closed string states to those of the black hole, the excitation levels of the closed string become
\begin{align}
N_{L}\sim \alpha' (M_{BH}^2-(q_p-q_w)^2), \label{eq:nlblack} \\
N_{R}\sim \alpha' (M_{BH}^2-(q_p+q_w)^2). \label{eq:nrblack}
\end{align}
Inserting (\ref{eq:nlblack}) and (\ref{eq:nrblack}) into the formula of the entropy of the closed string states (\ref{eq:stringentropy}), one finds
\begin{align}
S\sim \sqrt{N_R}+\sqrt{N_L}\sim l_s\frac{r_H^{d-3}}{G_N}\cosh \alpha_p \cosh \alpha_w. \label{eq:entropyofstringblackhole}
\end{align}
Thus, at the correspondence point (\ref{eq:correspondencepoint}), the entropy of the closed string states becomes of the same order as the Bekenstein-Hawking entropy (\ref{eq:blackentropy}).

Let us comment that the gravitational redshift does not affect the determinations of $N_L$ and $N_R$. At the correspondence point, the factor $(1-k(r))$ in the black hole solution (\ref{eq:ns2charge}) gets smeared out to order unity by the stringy corrections at $r\sim r_H$  \cite{Sen:1995in, Horowitz:1996nw, Peet:1997es}. Thus, the redshift factor near the horizon is of the order of $(\cosh^2 \alpha_p \cosh^2\alpha_w)$, which is very large when $\alpha_p$ and $\alpha_w$ are large. To see that this redshift  factor does not have an effect on (\ref{eq:nlblack}) and (\ref{eq:nrblack}), we consider the simple case $\alpha_p=0$.
In this case, $N_L$ is equal to $N_R$. In the near extremal limit $r_H/G_N\to 0, \alpha_w\to \infty$ keeping  $r_H^{d-3}e^{2\alpha_w}/G_N$  fixed, (\ref{eq:nlblack}) or (\ref{eq:nrblack})
becomes
\begin{align}
N\sim  RW\bigg(M_{BH}-\frac{RW}{\alpha'}\bigg). \label{eq:detn}
\end{align}
Since the mass of the extremal black hole is $M_{BH}^{(ext)}=RW/\alpha'$, $\Delta E=M_{BH}-RW/\alpha'$ is the excess energy above the extremality. Although the excess energy near the horizon is redshifted by the factor $\cosh \alpha_w$, since the radius $R$ is contracted by the same factor near the horizon, (\ref{eq:detn}) does not change even if we take into account the redshift. If we turn on $\alpha_p$, this does not affect the determinations of $N_L$ and $N_R$ \cite{Horowitz:1996nw}.

Finally, we comment that the free string formulas which we have used are valid if the string is highly excited \cite{Peet:1997es}. Using (\ref{eq:entropyofstringblackhole}) and $G_N\sim g_s^2l_s^{d-2}$ at the correspondence point, one finds
\begin{align}
g_s\sim \frac{(\cosh \alpha_p \cosh \alpha_w)^{\frac{1}{2}}}{(\sqrt{N_L}+\sqrt{N_R})^{\frac{1}{2}}}.
\end{align}
The local string coupling is given by $e^{\Phi(r)}g_s$. Thus, near the horizon, the local string coupling becomes
\begin{align}
e^{\Phi(r_H)}g_s\sim \frac{1}{(\sqrt{N_L}+\sqrt{N_R})^{\frac{1}{2}}},
\end{align}
which is much smaller than $1$ if the string is highly excited. Therefore, the free string formulas are valid near the horizon.

\subsection{Closed string transport coefficients at the correspondence point}
I discuss the correspondence between the closed string transport coefficients and the membrane transport coefficients. 
At first, one has to consider the mass dimensions of both the transport coefficients. The mass dimensions of the closed string transport coefficients are $d-1$. On the other hand, the mass dimensions of the membrane transport coefficients are $d-2$. This discrepancy comes from the fact that the physical degrees of freedom of the black hole seem to live on the horizon for a distant observer. In fact, the distant observer can not see inside the black hole.
Thus, I assume that at the correspondence point, the highly excited closed string states are on the stretched horizon of the black hole \cite{Susskind:1993ws, Sasai:2010pz}. This idea has been originally proposed in  \cite{Susskind:1993ws} to explain a macroscopic black hole entropy from highly excited string states.

The most convenient way to realize the string states on the stretched horizon is to reduce the longitudinal direction because in the light-cone gauge, the physical degrees of freedom of the string are given by the transverse oscillators $\alpha_n^i$. By a dimensional reduction along the $x^-$ direction, $V_{d-1}$ in the expressions of the closed string transport coefficients is replaced by $V_{d-2}\sim L^{d-2}$ \cite{Sasai:2010pz}. Then, the mass dimensions of the longitudinally reduced closed string transport coefficients are $d-2$, which match with those of the membrane transport coefficients.

Next, I compare the longitudinally reduced closed string transport coefficients with the membrane transport coefficients according to the black hole correspondence principle.

The longitudinally reduced closed string transport coefficients in the trivial background fields take the following form:
\begin{align}
\xitilde \sim \frac{1}{V_{d-2}}(\sqrt{N_{R}}+\sqrt{N_L}). \label{eq:xitilde}
\end{align}
Since the closed string states are assumed to live on the stretched horizon, $V_{d-2}$ is the area of the stretched horizon. Thus, from (\ref{eq:area}), one finds
\begin{align}
V_{d-2}\sim r_H^{d-2}\cosh \alpha_p \cosh \alpha_w. \label{eq:vd-2}
\end{align}
When the mass and charges of the closed string states are equal to those of the black hole, $\sqrt{N_R}+\sqrt{N_L}$ is given by (\ref{eq:entropyofstringblackhole}). Thus, at the correspondence point (\ref{eq:correspondencepoint}), $\xitilde$ becomes
\begin{align} 
\xitilde \sim \frac{1}{G_N}.
\end{align}

One might suspect that the hydrodynamical description of the string states is  valid at the correspondence point because the string size seems to be comparable to the ``microscopic scale" of the system.  To clarify this, let us find what the microscopic scale of the system is at the correspondence point. From (\ref{eq:entropyofstringblackhole}) and (\ref{eq:vd-2}), the entropy density of the string states on the horizon at the correspondence point is
\begin{align}
s= \frac{S}{V_{d-2}}\sim \frac{1}{G_N}=\frac{1}{l_p^{d-2}},
\end{align}
where $l_p$ is the Planck length. This means that there is one string state (or a string bit\footnote{In the neutral case, the total length of a highly excited string is  $\sim l_s\sqrt{N}$. Thus, the number of the string bits is $\sim \sqrt{N}$, which is of the same order as the entropy.}) per Planck unit area. Thus, the microscopic scale of the system is  $l_p$, not $l_s$. Since the area of the horizon $V_{d-2}$ is much bigger than the Planck unit area $l_p^{d-2}$ if the excitation levels $N_L$ and $N_R$ are much bigger than 1,   the hydrodynamics of  the string states can be defined at the correspondence point.
Probably, a string bit corresponds to an elementary particle of the fictitious fluid of the membrane paradigm.

The gravitational redshift does not have an effect on the closed string transport coefficients because the excitation levels of the string in (\ref{eq:xitilde}) do not depend on the redshift factor as was explained in the previous subsection. 
In that sense, the  calculations in string theory around the flat background spacetime seem to be  valid.

Thus,  the longitudinally reduced closed string transport coefficients become of the same order as the membrane transport coefficients except for the dilaton factors in (\ref{eq:aconductivity}), (\ref{eq:bconductivity}) and (\ref{eq:btrans}). These mismatches occur because  I have not considered the background dependence of the dilaton field near the horizon in the string calculation.
Let us take into account for the background dependence of the dilaton field.
The value of the dilaton field at the stretched horizon is given by
\begin{align}
e^{-4\Phi(r_H)}= \cosh^2 \alpha_p \cosh^2 \alpha_w.
\end{align}
Expanding the dilaton field around this value and the other fields around the trivial background, the action (\ref{eq:worldsheet2}) becomes
\begin{align}
S&\simeq S_0+S_1, \\
S_0&=\frac{1}{4\pi \alpha'}\int d^2\sigma e^{\frac{4\Phi(r_H)}{d-2}}\eta_{\mu\nu}(\dot{X}^{\mu}\dot{X}^{\nu}-X^{'\mu}X^{'\nu})+\delta_{ab}(\dot{X}^{a}\dot{X}^{b}-X^{'a}X^{'b}), \\
S_1&=\frac{1}{4\pi \alpha'}\int d^2\sigma [e^{\frac{4\Phi(r_H)}{d-2}}h_{\mu\nu}(X)P^{\mu\nu}+\frac{4}{d-2}e^{\frac{4\Phi(r_H)}{d-2}}\Phi(X) \eta_{\mu\nu}P^{\mu\nu}+\Gtilde_{ab}(X)P^{ab} \notag \\
&~~~~~~~~~~~~~~~~~+B_{ab}(X)Q^{ab} +2A_{a\mu}(X)P^{a\mu}+2B_{\mu a}(X)Q^{\mu a}+B_{\mu\nu}(X)Q^{\mu\nu}].
\end{align}
To obtain the standard action (\ref{eq:standard}), we rescale $X^{\mu}$ to $\exp(-\frac{2\Phi(r_H)}{d-2})X^{\mu}$. Then, $S_1$ becomes
\begin{align}
S_1&=\frac{1}{4\pi \alpha'}\int d^2\sigma [h_{\mu\nu}(X)P^{\mu\nu}+\frac{4}{d-2}\Phi(X) \eta_{\mu\nu}P^{\mu\nu}+\Gtilde_{ab}(X)P^{ab} \notag \\
&+B_{ab}(X)Q^{ab} +2A_{a\mu}(X)e^{-\frac{2\Phi(r_H)}{d-2}}P^{a\mu}+2B_{\mu a}(X)e^{-\frac{2\Phi(r_H)}{d-2}}Q^{\mu a}+B_{\mu\nu}(X)e^{-\frac{4\Phi(r_H)}{d-2}}Q^{\mu\nu}].
\end{align}
Therefore, one can obtain the closed string transport coefficients in the presence of the background dilaton field if one replaces
\begin{align}
P^{a\mu}&\to e^{-\frac{2\Phi(r_H)}{d-2}}P^{a\mu}, \\
Q^{a\mu}&\to e^{-\frac{2\Phi(r_H)}{d-2}}Q^{a\mu}, \\
Q^{\mu\nu}&\to e^{-\frac{4\Phi(r_H)}{d-2}}Q^{\mu\nu},
\end{align}
in the calculations of section  \ref{sec:cal}. Since the closed string transport coefficients are given by the commutators of $P^{MN}$ or $Q^{MN}$, the closed string transport coefficients induced by $A_{a\mu}, B_{a\mu}$ and $B_{\mu\nu}$ become
\begin{align}
\tilde{\xi}_{A_i^a}&\sim e^{-\frac{4\Phi(r_H)}{d-2}}\frac{1}{G_N}, \\
\tilde{\xi}_{B_{ia}}&\sim e^{-\frac{4\Phi(r_H)}{d-2}}\frac{1}{G_N}, \\
\tilde{\xi}_{B_{ij}}&\sim e^{-\frac{8\Phi(r_H)}{d-2}}\frac{1}{G_N},
\end{align}
at the correspondence point. These dilaton factors are the same as those in (\ref{eq:aconductivity}), (\ref{eq:bconductivity}) and (\ref{eq:btrans}). Thus, except for the bulk viscosity,  the closed string transport coefficients agree with the membrane transport coefficients at the correspondence point.

If one defines the dimensionless quantities by dividing the transport coefficients by the entropy densities, it is much simpler to discuss the correspondence. In addition, one does not need any numerical approximations. The dimensionless membrane transport coefficients are
\begin{align}
&\frac{\eta_{BH}}{s_{BH}}=\frac{d-2}{8}\frac{\chi_{\Phi}}{s_{BH}}=\frac{\chi_{G_{ab}}}{s_{BH}}=\frac{\chi_{B_{ab}}}{s_{BH}}=\frac{1}{4\pi}, \\
&\frac{\chi_{A^a_{i}}}{s_{BH}}=\frac{\chi_{B_{i a}}}{s_{BH}}=\frac{1}{4\pi}e^{-\frac{4\Phi(r_H)}{d-2}}, \label{eq:dimlesscond} \\
&\frac{\chi_{B_{ij}}}{s_{BH}}=\frac{1}{4\pi}e^{-\frac{8\Phi(r_H)}{d-2}}. \label{eq:dimlessbtrans}
\end{align}
On the other hand, the dimensionless closed string transport coefficients are
\begin{align}
&\frac{\eta}{s}=\frac{d-2}{8}\frac{\xi_{\Phi}}{s}=\frac{\xi_{G_{ab}}}{s}=\frac{\xi_{B_{ab}}}{s}=\frac{3}{2\pi c}, \\
&\frac{\xi_{A^a_{i}}}{s}=\frac{\xi_{B_{i a}}}{s}=\frac{3}{2\pi c}e^{-\frac{4\Phi(r_H)}{d-2}}, \\
&\frac{\xi_{B_{ij}}}{s}=\frac{3}{2\pi c}e^{-\frac{8\Phi(r_H)}{d-2}}. \label{eq:dimlessstring}
\end{align}
Thus, if $c=6$, both the dimensionless transport coefficients are exactly equal except for the bulk viscosity.
Although $c=6$ is not the central charge of the critical string theory, this value has been discussed to reproduced the correct numerical coefficient of the Bekenstein-Hawking entropy of the Schwarzshild black hole from string theory \cite{Halyo:1996xe}.

\section{Summary and comments} 
I have calculated the membrane transport coefficients and the closed string transport coefficients in the $D$ dimensional closed string backgrounds whose $D-d$ spatial directions are  toroidally compactified. There are two reasons why I have considered the toroidal compactification.  The first reason is to compare  the membrane transport coefficients with  the closed string transport coefficients in more general setting. The second reason is to introduce the Kaluza-Klein number $K$ and winding number $W$ in the closed string states. 

Then, I have discussed the correspondence between the membrane transport coefficients and the closed string transport coefficients according to the black hole correspondence principle.
 I have found that except for the bulk viscosity, the membrane transport coefficients are of the same order as the transport coefficients of the closed string states on the stretched horizon at the correspondence point even when one considers nonvanishing charges.  The nonvanishing dilaton factors of the membrane transport coefficients which appear in  (\ref{eq:aconductivity}), (\ref{eq:bconductivity}) and (\ref{eq:btrans}) have been reproduced from the string calculation if one takes into account for the background dilaton field near the horizon. Finally, I have shown that if the central charge $c$ is $6$, both the dimensionless transport coefficients are exactly equal except for the bulk viscosity.

Three comments are in order. Firstly, the value of the central charge $c=6$ might be explained if we assume type IIB theory compactified on $T^5$ \cite{Mathur:1997wb}.  If we increase the string coupling, the excitations of NS5-$\overline{\mathrm{NS5}}$ pairs  wraped on $T^5$ can occur because the mass of NS5-brane is proportional to $g_s^{-2}$. It is known that at a larger string coupling than the correspondence point, the excitations of NS5-$\overline{\mathrm{NS5}}$ pairs attached to the  string become entropically more favorable than the usual fundamental string excitations \cite{Mathur:1997wb}. By taking the dualities, this system becomes D1-D5-P system. Since the central charge of the effective string model of D1-D5-P system is 6, we will find the exact correspondence between the membrane paradigm and string theory. I will discuss this problem in my forthcoming paper.

Secondly, we can not find the correspondence between the black hole  and the fundamental string states concerning to the bulk viscosity  as it has been already reported in our previous paper \cite{Sasai:2010pz}. This is because the bulk viscosity of the membrane paradigm is negative $\zeta_{BH}=-\frac{1}{16\pi G_N}$ \cite{Thorne:1986iy}, while that of the fundamental string states is positive.
 The calculation of the bulk viscosity of the fundamental string states is given in the Appendix \ref{app}. It is still mysterious  why this mismatch occurs.

Finally, it might be interesting to see the correspondence between the fundamental string states with the fuzzball solutions \cite{Mathur:2005zp}. If one  finds how to obtain the transport coefficients of the fuzzball, one might be able to compare them with the closed string transport coefficients without reducing the longitudinal direction. Performing dualities to obtain D1-D5 system is also interesting.

\section*{Acknowledgments}
I would like to thank to Tohru Eguchi, Esko Keski-Vakkuri, Hiroshi Kunitomo, Yasuyuki Hatsuda,  Toshihiro Matsuo, Koichi Murakami, Yasuhiro Sekino, Fumihiko Sugino and Ali Zahabi for useful discussions and comments.  I also would like to thank to the participants of the workshop ``Towards New Developments in Field and String Theories" held at RIKEN on 17-19 December 2010 for variable discussions.
I was supported in part by JSPS-Academy of Finland bilateral scientist exchange programme.

\appendix

\section{Bulk viscosity of fundamental string states} \label{app}
I calculate the bulk viscosity of the fundamental string states by using the linear response theory. A bulk viscosity  in $d$ dimensional spacetime is defined by \cite{Gubser:2008sz}
\begin{align}
\delta T^{i}{}_{i}(\omega)=\frac{d-1}{2}i\omega \zeta h^{i}{}_{i}(\omega),
\end{align}
where $i=1,\cdots, d-1$ runs over all spatial components. Since in the light-cone coordinates, the expression of the bulk viscosity is not simple, I identify $\tau$ with $X^0$ instead of (\ref{eq:lightconegauge}). Therefore,
\begin{align}
X^0=2\alpha' M\tau.
\end{align}
Although we overcount the physical oscillations of the string by one in this setting,  we do not mind the numerical problem because what I would like to show  here is that the bulk viscosity of the fundamental string is positive. 

By the Kubo's formula (\ref{eq:kubo}), the bulk viscosity is
\begin{align}
\zeta&=\frac{2}{d-1}\lim_{\omega \to 0}\frac{V_{d-1}}{\omega}\delta^{ij}\delta^{kl}f_{\Tcalbar_{ij}\Tcalbar_{kl}}(\omega) \notag \\
&=\frac{2}{d-1}\frac{V_{d-1}}{2}\bigg(\frac{1}{4\pi \alpha^{'2}MV_{d-1}}\bigg)^2\delta_{ij}\delta_{kl}\delta_+^{ij,kl}\lim_{\omega \to 0}\frac{I(\omega)}{\omega} \notag \\
&=\frac{2}{V_{d-1}}\sqrt{\frac{6}{c}}(\sqrt{N_R}+\sqrt{N_L}).
\end{align}
Therefore, the bulk viscosity of the fundamental string is positive.
At the correspondence point, the bulk viscosity of the longitudinally reduced string  becomes
\begin{align}
\tilde{\zeta}\sim \frac{1}{G_N}.
\end{align}

On the other hand,  the bulk viscosity in the membrane paradigm is \cite{Thorne:1986iy}
\begin{align}
\zeta_{BH}=-\frac{1}{16\pi G_N}.
\end{align}
Therefore, we find the discrepancy between the bulk viscosity of the membrane paradigm and that of the fundamental string states  although both the absolute values  are of the same order.


\begin{thebibliography}{20}
\bibitem{Susskind:1993ws}
  L.~Susskind,
  ``Some speculations about black hole entropy in string theory,''
  arXiv:hep-th/9309145.


\bibitem{Sen:1995in}
  A.~Sen,
  ``Extremal black holes and elementary string states,''
  Mod.\ Phys.\ Lett.\  A {\bf 10}, 2081 (1995)
  [arXiv:hep-th/9504147].


\bibitem{Horowitz:1996nw}
  G.~T.~Horowitz and J.~Polchinski,
  ``A correspondence principle for black holes and strings,''
  Phys.\ Rev.\  D {\bf 55}, 6189 (1997)
  [arXiv:hep-th/9612146].
  
  

 
  

\bibitem{Strominger:1996sh}
  A.~Strominger and C.~Vafa,
  ``Microscopic Origin of the Bekenstein-Hawking Entropy,''
  Phys.\ Lett.\  B {\bf 379}, 99 (1996)
  [arXiv:hep-th/9601029].

\bibitem{Callan:1996dv}
  C.~G.~Callan and J.~M.~Maldacena,
  ``D-brane Approach to Black Hole Quantum Mechanics,''
  Nucl.\ Phys.\  B {\bf 472}, 591 (1996)
  [arXiv:hep-th/9602043].
  






\bibitem{Sasai:2010pz}
  Y.~Sasai, A.~Zahabi,
  ``Shear viscosity of a highly excited string and the black hole membrane paradigm,''
  Phys.\ Rev.\  {\bf D83}, 026002 (2011).
  [arXiv:1010.5380 [hep-th]].
  


\bibitem{Damour:1978cg}
  T.~Damour,
  ``Black Hole Eddy Currents,''
  Phys.\ Rev.\  D {\bf 18}, 3598 (1978).


\bibitem{Znajek:1978}
R.~L.~Znajek,
``The electric and magnetic conductivity of a Kerr hole,"
Mon.\ Not.\ R.\ Astron.\ Soc.\ {\bf 185}, 833 (1978).





\bibitem{Thorne:1986iy}
  K.~S.~Thorne, R.~H.~Price and D.~A.~Macdonald,
  ``BLACK HOLES: THE MEMBRANE PARADIGM,''
{\it  NEW HAVEN, USA: YALE UNIV. PR. (1986) 367p}





\bibitem{Parikh:1997ma}
  M.~Parikh and F.~Wilczek,
  ``An action for black hole membranes,''
  Phys.\ Rev.\  D {\bf 58}, 064011 (1998)
  [arXiv:gr-qc/9712077].


\bibitem{Polchinski:1998}
J.~Polchinski,
  ``String theory. Vol. 1: An introduction to the bosonic string,''
{\it  Cambridge, UK: Univ. Pr. (1998) 402 p}.



\bibitem{Horowitz:1992jp}
  G.~T.~Horowitz,
  ``The dark side of string theory: Black holes and black strings.,''
  In *Trieste 1992, Proceedings, String theory and quantum gravity '92*, World Scientific Pub. Co. Inc. (1993) pp.55-99.
  [hep-th/9210119].

\bibitem{Peet:1997es}
  A.~W.~Peet,
  ``The Bekenstein formula and string theory (N-brane theory),''
  Class.\ Quant.\ Grav.\  {\bf 15}, 3291-3338 (1998).
  [hep-th/9712253].



\bibitem{Peet:2000hn}
  A.~W.~Peet,
  ``TASI lectures on black holes in string theory,''
  
  [hep-th/0008241].

\bibitem{Iqbal:2008by}
  N.~Iqbal and H.~Liu,
  ``Universality of the hydrodynamic limit in AdS/CFT and the membrane
  paradigm,''
  Phys.\ Rev.\  D {\bf 79}, 025023 (2009)
  [arXiv:0809.3808 [hep-th]].
  


\bibitem{Halyo:1996xe}
  E.~Halyo, B.~Kol, A.~Rajaraman, L.~Susskind,
  ``Counting Schwarzschild and charged black holes,''
  Phys.\ Lett.\  {\bf B401}, 15-20 (1997).
  [hep-th/9609075].



\bibitem{Damour:1999aw}
  T.~Damour, G.~Veneziano,
  ``Selfgravitating fundamental strings and black holes,''
  Nucl.\ Phys.\  {\bf B568}, 93-119 (2000).
  [hep-th/9907030].
 
\bibitem{Landau}
L.~D.~Landau and E.~M.~Lifshitz,
``Statistical Physics, 3rd Edition, Part 1,"
{\it Butterworth-Heinemann (1980) 544 p}.
 
 
\bibitem{Chaikin:book}
P.~M.~Chaikin and T.~C.~Lubensky,
``Principles of condensed matter physics,"
{\it Cambridge, UK: Cambridge Univ. Pr. (2000) 699 p}. 


\bibitem{Mathur:1997wb}
  S.~D.~Mathur,
  ``Emission rates, the correspondence principle and the information paradox,''
  Nucl.\ Phys.\  {\bf B529}, 295-320 (1998).
  [hep-th/9706151].



\bibitem{Mathur:2005zp}
  S.~D.~Mathur,
  ``The Fuzzball proposal for black holes: An Elementary review,''
  Fortsch.\ Phys.\  {\bf 53}, 793-827 (2005).
  [hep-th/0502050].


\bibitem{Gubser:2008sz}
  S.~S.~Gubser, S.~S.~Pufu, F.~D.~Rocha,
  ``Bulk viscosity of strongly coupled plasmas with holographic duals,''
  JHEP {\bf 0808}, 085 (2008).
  [arXiv:0806.0407 [hep-th]].

\end{thebibliography}
\end{document}